\documentclass[aps,pra,twocolumn,showpacs,groupedaddress]{revtex4-1} 
\usepackage{graphicx}  
\usepackage{mathrsfs, amssymb}   
\usepackage{subfigure}
\usepackage{epstopdf}
\begin{document}
\title{Quantitative measurement of orbital angular momentum in electron microscopy}

\author{L. Clark*, A. B\'ech\'e, G. Guzzinati, J. Verbeeck}
\affiliation{EMAT, University of Antwerp, Groenenborgerlaan 171, 2020 Antwerp, Belgium}
\email{laura.clark@uantwerp.ac.be}

\begin{abstract}
Electron vortex beams have been predicted to enable atomic scale magnetic information measurement, via transfer of orbital angular momentum.
Research so far has focussed on developing production techniques and applications of these beams.
However, methods to measure the outgoing orbital angular momentum distribution are also a crucial requirement towards this goal.
Here, we use a method to obtain the orbital angular momentum decomposition of an electron beam, using a multi-pinhole interferometer.
We demonstrate both its ability to accurately measure orbital angular momentum distribution, and its experimental limitations when used in a transmission electron microscope.
\end{abstract}

\maketitle

\section{Introduction}
That light waves can carry an orbital angular momentum (OAM), in addition to (and distinct from) their spin angular momentum, has now been known for over twenty years \cite{Allen1992}. Waves carrying OAM are typically exemplified by vortex beams, with an azimuthally varying phase \cite{OptAngMomBook}.
A substantial amount of research has been performed into the fundamental properties of such beams, as well as into methods of production, OAM measurement, and a vast array of applications \cite{BabikerNewBook, allenpadgettbabiker, YaoOAM}. Such studies now form the field of research known as singular optics \cite{singopt}.
Recently, this knowledge has also been applied in the field of electron microscopy to create electron vortex beams, also carrying an OAM,  with much hope for opening new information transfer possibilities (for example, about magnetic samples) \cite{Bliokh2007, VerbeeckNature}. Towards this aim, we have been studying methods of OAM measurement that are applicable within a typical transmission electron microscope (TEM).

Several methods of OAM measurement have been developed over the past few years in light optics \cite{laverymeasurement, vasnetsovobservation, Berkhoutmeasurement}. Here we discuss in particular the multi--pinhole interferometer (MPI)~\cite{BerkhoutAnisotropy}, as it can be installed in a TEM without major modification, making it a good candidate for such preliminary studies.
The very first steps towards methods of OAM measurement in the TEM have been developed by our group and others \cite{VerbeeckNature, Giulio2014, SaitohMeasuring}. However, these methods so far are limited to the cases where OAM and topological charge are equivalent (thus, only applicable when the intensity is cylindrically symmetric) \cite{opticalcurrents}, or require time-consuming holography and wavefront reconstruction techniques \cite{bechemagnetic}. Here we seek to loosen this restriction, enabling a measurement of the relative weight of a number of azimuthal modes present in a wavefront, about a chosen axis, which greatly enhances the potential for application in experiments.

In this article, we both demonstrate the application of an MPI within a TEM, and discuss the suitable applications and limitations of such a tool. 

\section{Phase and intensity retrieval using a multi-pinhole interferometer}
\begin{figure}
 \centering
 \includegraphics[width=1.0\linewidth]{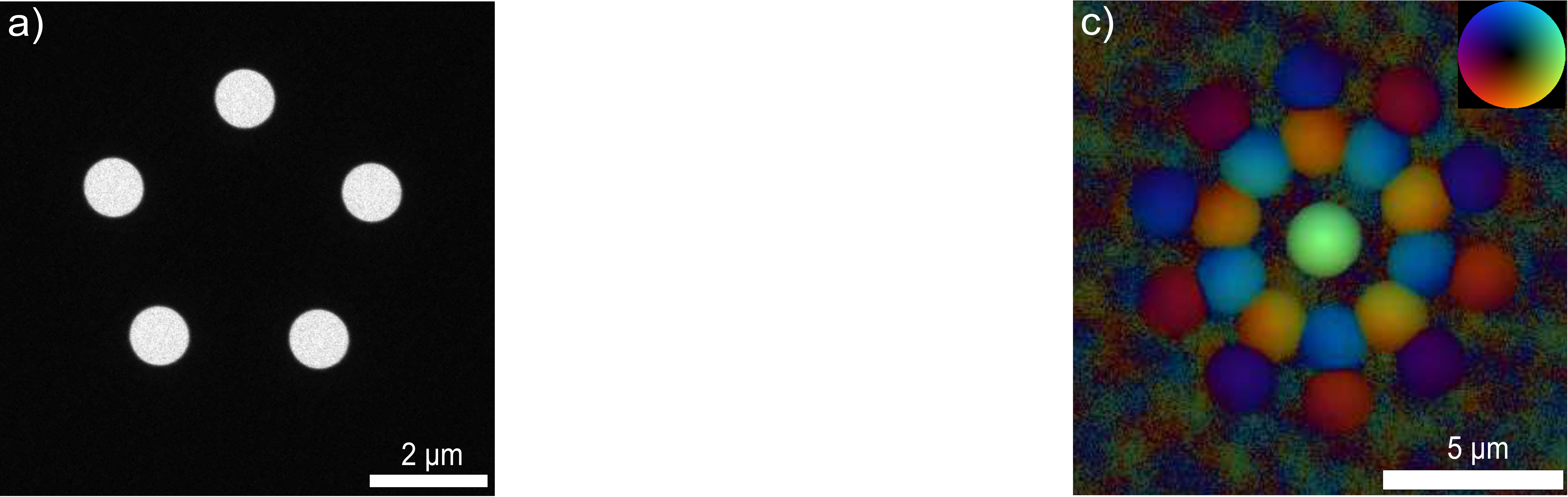}

 \caption{Demonstration of the key experimental components, in the $\ell=-1$ case. (a) TEM image of MPI in the selected area plane, (b) Diffraction pattern recorded from centred $\ell=-1$ beam impinging on the MPI, (c) Inverse Fourier transform of the recorded $\ell=-1$ diffraction pattern where brightness represents intensity, and hue represents phase from $0$-$2\pi$.  \label{MPI1}}
\end{figure}

As a general term, an MPI refers to any interferometer comprising of a set of small holes \cite{shicharacterizing, limeasuring, li2010measuring}. However, here we consider a particular subset, in which the $n$-pinholes are evenly spaced on a circle, as illustrated in Figure \ref{MPI1}a. 

This category of MPIs is of interest as it permits both a trivial qualitative topological charge measurement \cite{BerkhoutPRL}, and importantly, a simple quantitative analysis of the approximate OAM spectrum of an input wavefront \cite{Guo}.

To measure the OAM spectrum from a region of a beam, we place the MPI into the path of the OAM carrying beam, of which a typical example is a vortex beam described as:
\begin{equation}
\Psi=~A(r)\exp(\imath \ell \phi)\exp(\imath k_z z),
\end{equation}
where $(r,\phi,z)$ are the cylindrical coordinates, $k_z$ the forward momentum of the wave, and $\ell$ an integer known as the winding number, determining how much OAM is carried per electron \cite{OptAngMomBook}. 

\begin{figure}
 \centering
 \includegraphics[width=0.9\linewidth]{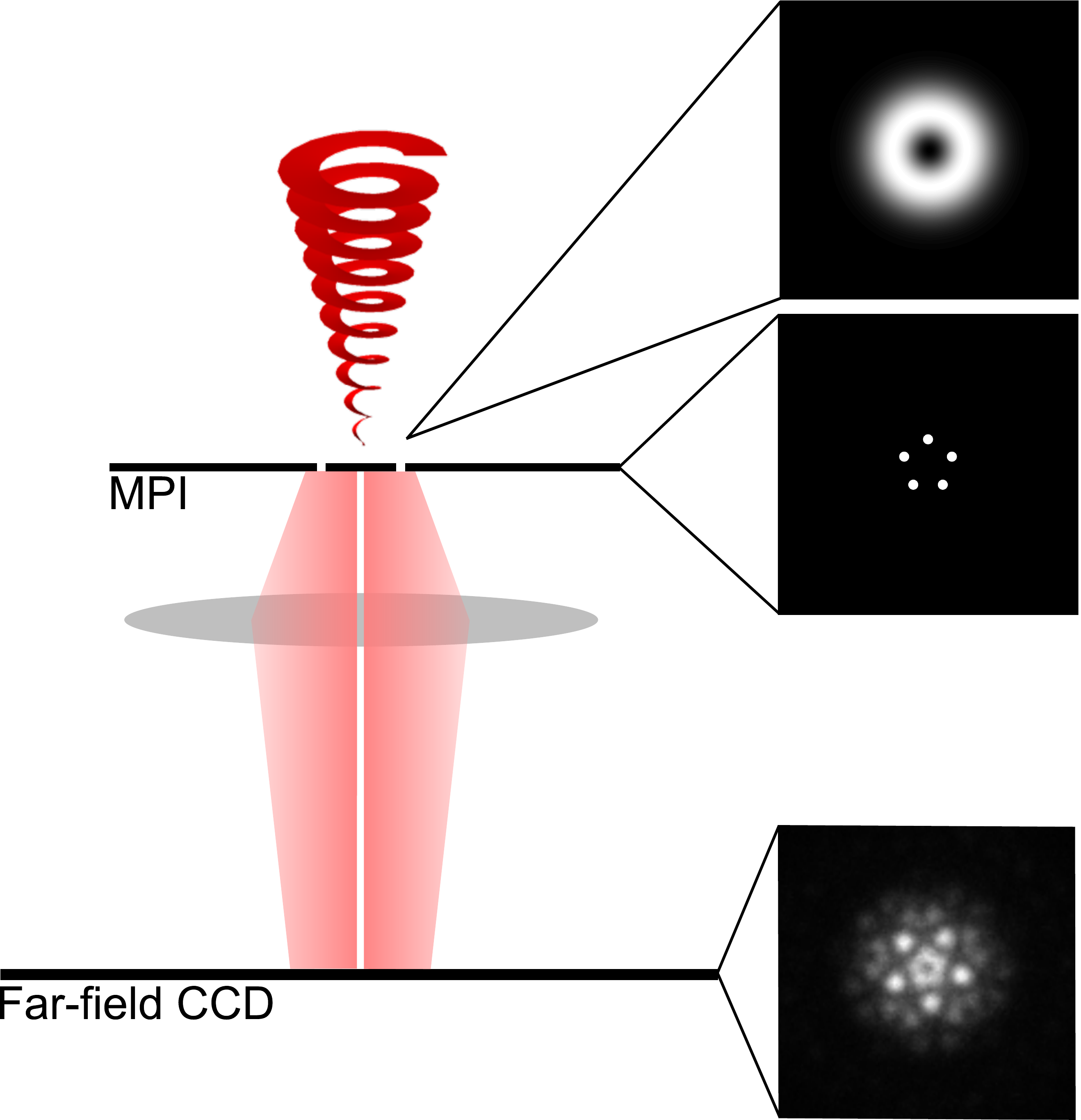}
 \caption{Diagram of experimental setup, showing vortex beam propagation and diffraction. \label{initialsetup}}
\end{figure}

We then record the intensity of the far-field diffraction pattern of this set-up, an example of which is illustrated in Figure \ref{MPI1}b.  
This diffraction pattern alone can differentiate between different orders of pure vortex states \cite{BerkhoutPRL}, but cannot directly determine relative mode weightings from a mixed vortex state.
To achieve this, we follow the routine of Guo \emph{et al.} \cite{Guo}, and expand on their discussion to clarify some details.
The experimental setup is illustrated in figure \ref{initialsetup}.

Firstly, the autocorrelation function of the MPI plane must be calculated. This can be done by taking advantage of the Wiener-Khinchin theorem \cite{chatfield}:
\begin{equation}
\mathscr{A}(\Psi)=\mathscr{F}^{-1}(|\mathscr{F}(\Psi)|^2).
\end{equation}
Where, $\mathscr{A}$ represents autocorrelation, $\mathscr{F}$ represents Fourier transformation, and $\Psi$ is a $2D$ wavefunction, in the plane perpendicular to the propagation axis.
Note that $|\mathscr{F}(\Psi)|^2$ is precisely our recorded diffraction pattern.
Accordingly, $\mathscr{A}(\Psi)$ can be calculated by simply inverse Fourier transforming a centred experimental diffraction pattern.

Autocorrelation functions have a high value at positions representing the vector between highly similar features of the input array. It thus follows that peaks will be present in our autocorrelation function (as shown in Fig. \ref{MPI1}c) at displacements from the centre position, equivalent to the displacement vector between one pinhole and another, in addition to a central peak representing the summed self-overlap of all of the pinholes. Consequently, the intensity of the off-centre peaks is a product of the wave passing through two relevant, contributing pinholes, over an area proportional to the overlapping of two disc functions.
Furthermore, the value of the phase of such a peak is related to the phase difference between two contributing pinholes \cite{Guo}, provided that the peaks in the autocorrelation function do not overlap with one another, and that the pinholes are sufficently small that the wavefront can be considered as constant across each hole. 
Such undesirable overlapping can be avoided through careful design of the MPI, discussed in section \ref{exp1section}.
The relative intensity and phase of the wavefront at each of the pinhole positions can thus be reconstructed (aside from an overall phase factor). From this we can approximate the OAM make-up of the sampled region, using the usual spiral-harmonic decomposition techniques \cite{MolinaTerriza2002}.

\section{Experimental demonstration of the MPI}\label{exp1section}

In this paper, we demonstrate two experiments using the MPI with an electron wave.
Firstly, we show efficient measurement of centered electron vortex beams of orders $\ell=\{-1,0,1,2\}$, and find these to be in rather pure states, showing that the method is insensitive to the inherent experimental imperfections.

Secondly, we perform the same analysis, over a series of illumination conditions, gradually displacing vortex beams of order $\ell=\pm 1$, away from an exactly centred position on the MPI, until the bright ring is no longer illuminating the interferometer. 
This enables a deeper consideration of the limitations of the MPI technique for OAM quantification within transmission electron microscopy.
\subsection{Centred Electron Vortices}
\begin{figure}
\centering
\includegraphics[width=1.0\linewidth]{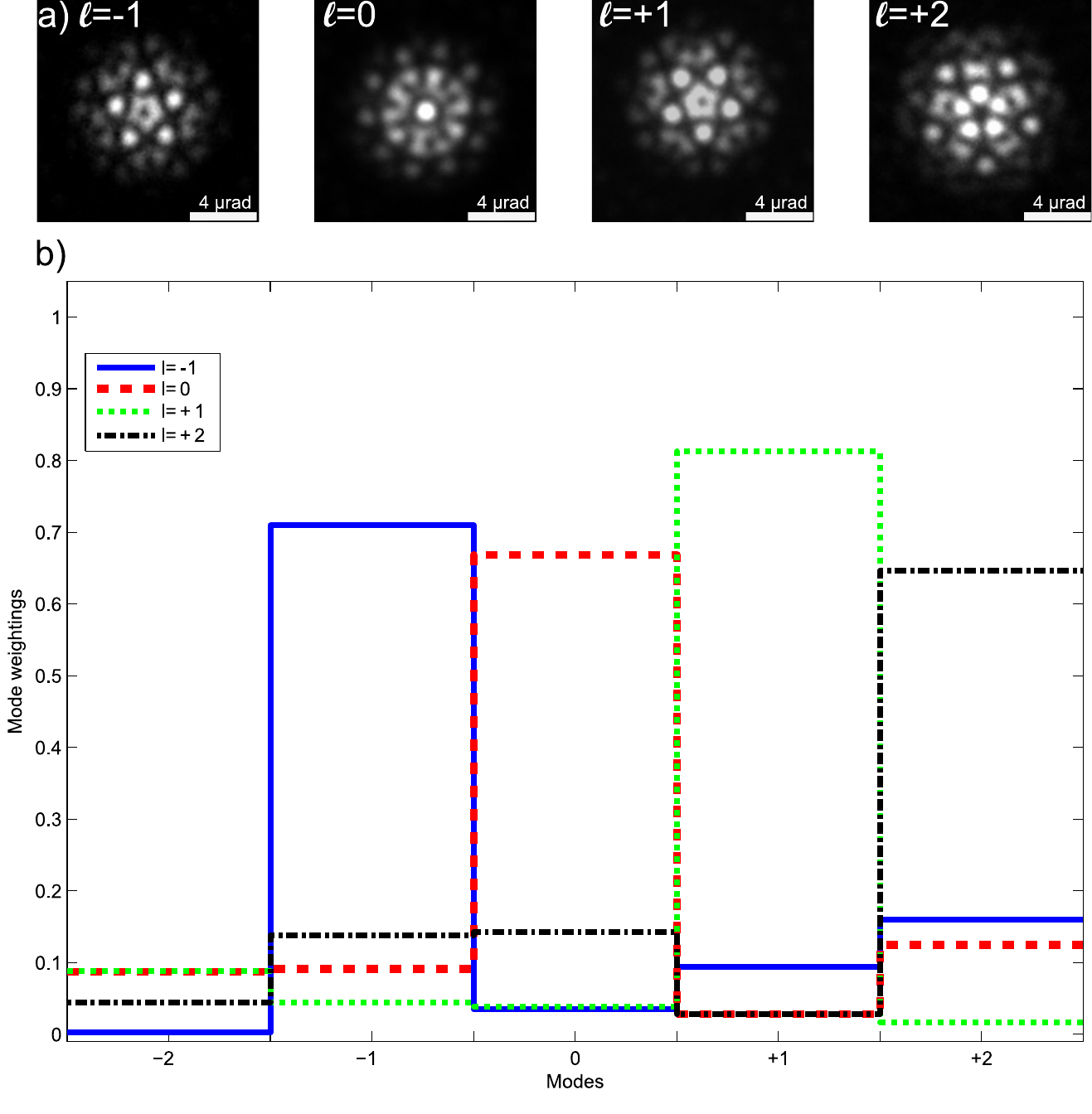}
\caption{a) Experimentally recorded diffraction patterns from the MPI, illuminated with a centred vortex beam of orders ${\ell=\{-1, 0, +1, +2\} }$. b) The resulting OAM mode decompositions from computational analysis of these diffraction patterns. \label{CentredVortedDecomps}}
\end{figure}
With an accelerating voltage of $200$kV (thus, a wavelength of $\lambda \approx 2.51$pm), we produce a set of vortex beams by illuminating a forked holographic mask in the condensor plane of the X-AntTEM, a probe aberration-corrected FEI Titan$^3$ TEM (procedure as described in Ref.\cite{VerbeeckNature}). This method of electron vortex beam production is chosen over other available techniques (such as the spiral holographic mask \cite{VerbeeckSTEM}, probe-aberration corrector misalignment \cite{Clark2013} or the newly demonstrated magnetic needle \cite{bechemagnetic, blackburnvortex}), as the forked-mask method produces vortex beams of various orders and high mode purity, permitting an initial verification of the MPI technique.

We defocus the beam slightly such that the MPI is maximally illuminated when the chosen vortex beam is centred. The vortex beams have a radius of approximately $35$~nm in the image plane, measured to the ring of highest intensity, to coincide with the centre of the demagnified MPI pinholes. We record the diffraction pattern of this set-up with a CCD camera, located in the far field.

To avoid intermixing of information during the analysis process from overlapping peaks in the autocorrelation image,  we require that $\pi (a - \frac{b}{2}) \ge n b$, where $a$ is the MPI radius (to the outer edge of the pinholes), $b$ the pinhole diameter and $n$ the number of pinholes.
To allow for some system imperfections (instabilities, residual aberrations etc.) we used $a=2.5\mu$m and $b=1\mu$m with an MPI of $n=5$ pinholes, as illustrated in Fig. \ref{MPI1}a, produced on an FEI Helios Dual-Beam Focussed Ion Beam instrument (FIB).
With this system, five distinct OAM modes can be measured, typically over $\ell=\{-\lfloor\frac{n}{2}\rfloor , \lfloor\frac{n}{2}\rfloor \}$ \cite{BerkhoutPRL}.
However, there is some freedom to choose which modes are to be measured by the MPI -- a factor which can be both an advantage and a disadvantage.
Higher order modes of $|\ell| > \frac{n}{2}$ are wrapped back around onto one of the measurable modes, leading to aliasing effects. This is a fundamental sampling limit of the system, requiring some consideration as to the appropriateness of the experiment to  which it is applied.
These points are considered in more detail in section \ref{sec:Discussion}.

We illuminate the MPI with centred $\ell=\{-1,0,1,2\}$ beams and record the respective diffraction patterns in the far field, displayed in Fig. \ref{CentredVortedDecomps}a, to demonstrate both chiralities, a higher order beam, and a non-vortex beam.

With this gathered data, we begin the analysis process. 
The experimental diffraction pattern images are firstly centred, and then inverse Fourier transformed to obtain the autocorrelation function of the input wavefunction.
 In each autocorrelation image, there are ten possible pentagon patterns formed by the peaks, from which the phase differences can be retrieved \cite{BerkhoutPRL}.
Recording the values of the peaks from any two of the pentagons, enables calculation of the relative phase and intensity at each of the pinholes, to produce an approximation to the OAM spectrum.

As shown in Fig. \ref{CentredVortedDecomps}b, we experimentally find a good approximation to this, with around $70\%$ of each wavefront sampled measured in the expected mode, and some spreading into the other modes. This mode spreading is a result of a number of causes, including residual aberrations, imperfect centering and limited coherence \cite{Vasnetsov, Clark2013}. However, we note that this mode spreading is limited, confirming that the electron vortex beams produced by holographic forked mask are a highly pure, and robust structure which can be reliably produced experimentally.

\subsection{Electron Vortex Shift Series}

\begin{figure}
\centering
\includegraphics[width=1\linewidth]{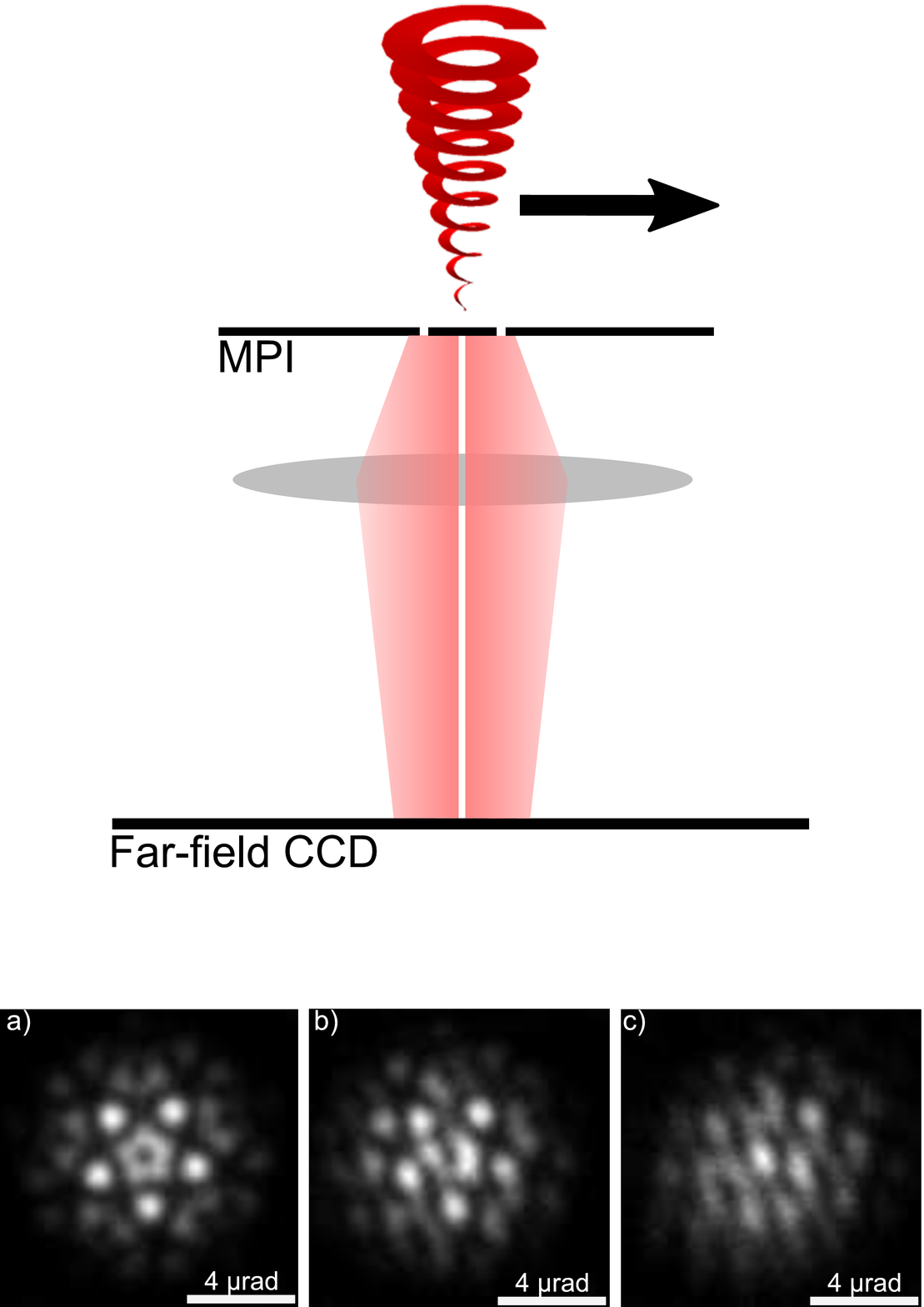}
 \caption{Diagram illustrating the shift series experiment (upper panel) and selected recorded diffraction patterns from an $\ell=+1$ vortex beam (lower panels). a) Shift of $6$nm, b) Shift of $30$nm, c) Shift of $42$nm. \label{shiftdiag}}
\end{figure}

Following the procedure as for the first experiment, we take an image of the centred diffraction pattern, and then shift the beam across the MPI, by a step of approximately $6$~nm and record the diffraction pattern again. We repeat this process, producing a series of diffraction patterns until the MPI is no longer illuminated by the high intensity vortex ring. We produced this series for both the $\ell=+1$ and $\ell=-1$ cases to allow for a consistency check, and process the data to find OAM mode decompositions at each position in the series. The set-up is illustrated in figure \ref{shiftdiag}.

\begin{figure*}
 \centering
 \includegraphics[width=0.75\linewidth]{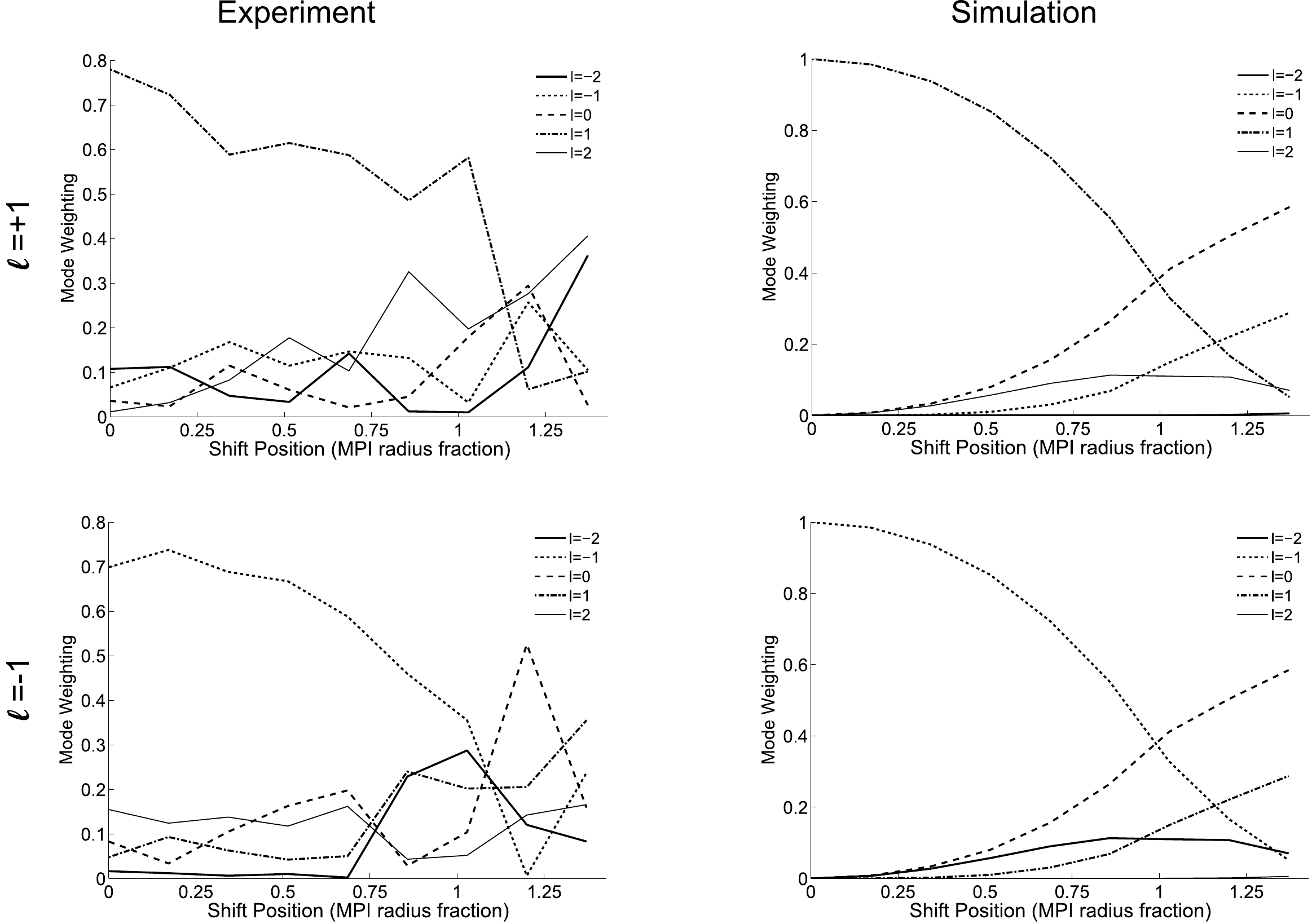}
 \caption{OAM spectra produced from an electron vortex beam, shifted away from the centre of the MPI, upper row, $\ell=+1$; lower row, $\ell=-1$; left hand side, experimental MPI data; right hand side, simulated MPI data.}
 \label{shiftFig}
\end{figure*}

At high values of shift, this analysis process fails, as not all pinholes remain well illuminated, leading to highly distorted diffraction patterns.  The progression of the pattern from the centred case, to off-centred is shown in figure \ref{shiftdiag}. While fig \ref{shiftdiag}a, still resembles the centred case, by the $30$nm shift of fig. \ref{shiftdiag}b, there is a notable change in the structure of the pattern. With a $42$nm shift (fig. \ref{shiftdiag}c), the $\ell=0$ pattern is beginning to form, albeit with a substantial amount of experimental noise due to the low illumination levels. This experimental noise means the required pentagon patterns cannot be identified in the resulting autocorrelation function, for the most shifted positions. Accordingly, from our data, we are only able to produce a mode decomposition for the first 9 recorded diffraction patterns in each series (over a total displacement of $36$nm). These are presented in fig \ref{shiftFig}, alongside results for a similar simulated MPI setup.

In an ideal system, we would expect the experimental plots to be identical to one another, other than the opposite handedness, as seen in the simulated decomposition plots. We note that the experimental $\ell=+1$ data is slightly shifted right with respect to the $\ell=-1$ data. This is likely due to limitations of the accuracy of the experimental shifting and positioning. Such an effect can be explained by a mis-positioning of the incident beam evaluated to $\approx 6$ nm, over the series, which is within the range of reasonable experimental error.

When the beams are centred, we replicate the diffraction pattern achieved in fig. \ref{MPI1}b. At small shifts from the centred position, the sharp peak in the OAM decomposition (as shown in fig. \ref{CentredVortedDecomps}) broadens slightly, as expected, following Vasnetsov \emph{et al.} \cite{Vasnetsov}, behaving in a highly predictable manner. As the shift increases, the intensity variations of the field also require consideration as trivial assumptions of a flat intensity profile are no longer appropriate. This changes the OAM composition, in a more complex, but still interpretable way. However, at the highest values of shift presented here, the data becomes less reliable, as a number of limiting factors are involved.
This limit of reliability of MPI OAM decomposition is discussed in detail in the following section.

\section{Discussion}\label{sec:Discussion}

Through these two experiments a number of interesting points to consider are raised. The MPI clearly has ability to measure the OAM composition of a field to some extent, but the limits themselves are not immediately clear. We shall begin by considering the measurement restrictions due to experimental limitations.

In electron microscopy, aberrations are time-varying, and can often be significant. We find here that the structure of the diffraction patterns is highly sensitive to aberrations. 
In particular, meticulous attention should be paid to minimising the diffraction astigmatism, to achieve good results from this set-up.

A second experimental challenge is that in the TEM, the relative orientation of the MPI and the diffraction patterns is unknown, due to Larmor rotation, and an unknown number of cross-overs in the microscope column. This means that the absolute handedness of a beam cannot be determined from this method alone, as five of the pentagons in the autocorrelation image are the complex conjugates of the other five available. To obtain this information, a separate calibration must be performed with a vortex of known chirality.

The pinholes of our MPI are placed on a circle of radius of $2.5 \mu$m. This size was chosen as at this scale, the MPI can be produced accurately and consistently on our FIB. Our field-sampling pinholes are all on the edges of this ring, and so sampling occurs over a circle with radius $\approx 35 \, nm$, when taking the demagnification due to projection into account. However, our wavelength is of the order of picometres, and our convergence angle is of the order of milliradians, so there may be field features of the same or smaller scale as the sampling frequency. If the variation of the field is more rapid than double the sampling frequency, an accurate measurement of the field cannot be guaranteed, as an angular analogue of the Nyquist theorem comes into effect. Such a situation is clearly possible in this setup. Accordingly, only areas of the field with variations in phase and intensity on a scale of tens of nanometres are appropriate, when producing OAM-mode decompositions with this method and set-up.

Sampling the electron wave at only $n$ points also results in a more fundamental limitation: aliasing.
All modes present in the sampled region may only be represented by one of $n$ options. Modes greater than the upper or lower limit of the MPI, are wrapped back around. Subsequently, modes of $(\ell+n)$ are recorded as $\ell$, and cannot be distinguished from one measurement. Options to improve this include using an MPI with a greater $n$, but this would require a larger radius overall, improving OAM measurement, but reducing spatial resolution, or, having multiple MPIs available with different $n$ values which would allow for a greater number of solutions by factorisation, but this would require a more complex setup.

Fields with significant weightings across more than $n$ modes cannot be faithfully measured by an MPI (for example, vortex beams with strong tilts or combinations of shift and tilt \cite{Vasnetsov}, or other substantial deviation from circular symmetry). However, non-canonical vortices, with only a few OAM components would still be expected to yield an MPI decomposition close to the true values.

Additionally, while the finite number of data points leads to an angular aliasing effect, with the pinholes placed at only one radial value, we are also limited in obtaining radial structure information. The phase and intensity variations around a circular path can differ with $r$. A recent example of this is shown in figure 5 of the work by Emile \emph{et al.} \cite{EmileOAM}. Such radial variations in OAM components cannot be measured using this simple MPI setup.

Furthermore, estimating the OAM of a region, from a sample of $n$-points on the perimeter of the region, means that small changes in intensity and phase near to a pinhole can have significant effects on the output spectrum. 
Uneven illumination of the pinholes has to be avoided as this strongly degrades the signal to noise ratio in the autocorrelation measurement.

\section{Conclusions}

The quantitative measurement of OAM spectra, as demonstrated here substantially increases the information we can retrieve from an electron beam.

The MPI technique is simple to employ experimentally, requiring only one new aperture, and careful alignment of the microscope, in strong contrast to time consuming holography or wavefront reconstruction techniques.
The analysis is straighforward, and we have demonstrated a good match between theory and experiment for a variety of input wavefronts, with the best outcomes from the evenly illuminated cases.

This work presents a real step towards the overarching goal to measure OAM spectra of electron vortex beams after having passed through a sample, by providing a proof of principle that such data can be obtained from a simple experimental setup.

\begin{acknowledgments}
L.C., A.B., G.G. and J.V. acknowledge funding from the European Research Council under the 7th Framework Program (FP7), ERC Starting Grant No. 278510 - VORTEX.
A.B. and J.V. also acknowledge financial support from the European Union under the 7th Framework Program (FP7) under contract for the Integrated Infrastructure Initiative No. 312483 ESTEEM2 and ERC grant No. 246791 - COUNTATOMS.
\end{acknowledgments}

\bibliography{OAMquantbib2}
\end{document}